\documentclass[11pt,fleqn]{article}
\usepackage{amsmath,amsfonts,amssymb,cite}
\usepackage{graphicx}
\usepackage{color}

\def\noi{\noindent}

\newcommand{\Title}[1]{\noi {{\Large\bf #1}}\\[1ex]}

\def\Aunames#1{\noi{\bf #1}}
\def\au#1{${}^{#1}$}
\def\Addresses#1{\medskip\noi \protect
	\begin{description}\itemsep -3pt {\it #1} \end{description}}
\def\adr#1#2{\item[${}^{#1}$]{\it #2}}

\def\email#1#2{\footnotetext[#1]{e-mail: #2}\addtocounter{footnote}{1}}

\def\Funding#1{\subsection*{Funding} #1}

\def\ConflictThey{\subsection*{Conflict of interest} 
	The authors declare that they have no conflicts of interest.}

\def\GR{general relativity}

\def\bhs{black holes}
\def\wh{wormhole}
\def\whs{wormholes}

\usepackage{color}

\begin{document}
\thispagestyle{empty}

\Title{Current problems and recent advances in wormhole physics\\[5pt]
		{\normalsize \it (Editorial for a Special issue of Universe journal)}}
		
\Aunames{Kirill A. Bronnikov,\au{a,b,c,1} Sergey V. Sushkov\au{d,2}}

\Addresses{\small
\adr a {Center of Gravitation and Fundamental Metrology, VNIIMS, 
		Ozyornaya ul. 46, Moscow 119361, Russia} 
\adr b  {Peoples' Friendship University of Russia (RUDN University), 
		6 Miklukho-Maklaya St, Moscow, 117198, Russia}
\adr c  {National Research Nuclear University ``MEPhI'', 
		Kashirskoe sh. 31, Moscow 115409, Russia}
\adr d {Institute of Physics, Kazan Federal University, 
		Kremliovskaya str. 16a, Kazan 420008, Russia}
		}

\bigskip
% ==============================================
\email 1 {kb20@yandex.ru}
\email 2 {sergey\_sushkov@mail.ru}

  Wormholes are hypothetical space-time tunnels with nontrivial topologies capable of connecting 
  either two distant regions of the same universe or two different universes \cite{MorTho, Visser, LoboRew}. 
  From the theoretical point of view, the possibility of their existence is problematic but cannot be ruled out.   
  If wormholes do exist, many unusual phenomena can be expected. Among them, probably the most
  exciting ones are shortcuts providing interstellar, intergalactic or inter-universe trips and even time
  machines. 

  At present, wormholes are extremely attractive and popular objects for research, it is sufficient to 
  mention that the word ``wormhole'' is found in the titles of 1614 articles on the resource arXiv.org for 
  all years and 175 articles for the last 12 months (as of 19.01.2023). Although many mathematical 
  and physical properties of these objects have been discovered and studied in the recent decades, 
  there remain multiple unsolved problems {and opportunities of interest to be explored.}
  
  The present Special Issue {\bf ``Recent Advances in Wormhole Physics''} is aimed at enlightening some 
  recent results in selected areas of wormhole physics. The issue is a collection of fourteen papers 
  \cite{1,2,3,4,5,6,7,8,9,10,11,12,13,14} {that fairly well characterizes the diversity of subjects
  and methods of \wh\ physics.  }
  
  Very roughly, \wh\ studies may be classified as follows:
\begin{itemize}
\item
	A search for \wh\ solutions in \GR\ and other theories of gravity, investigations of their 
	properties and conditions of their existence.
\item  
     Studies of mathematical, physical, metaphysical and philosophical consequences of possible
     \wh\ existence.  
\item
	Assuming that \whs\ do exist in the Universe, studies of their astronomical and astrophysical 
	manifestations, in particular, their possible observational distinctions from \bhs.
\end{itemize}
  Needless to say that many studies combine some or all of these trends. Nevertheless, it can be more or 
  less conditionally observed that the first trend is represented by the papers \cite {1,10,11,13}, the second 
  one by \cite{2,4,8,12}, and the third one by \cite{3,5,6,7,9,14}.

  Two of the papers are brief reviews \cite{1,6}.
  Thus, Takafumi Kokubu and Tomohiro Harada \cite{1} consider different physical aspects of 
  thin-shell \whs, {both in GR and in Einstein--Gauss--Bonnet gravity, including spherical, planar
  and hyperbolic symmetries of space-time and the stability issues.}
  Cosimo Bambi and Dejan Stojkovic \cite{6} describe astronomically observable effects of \whs\ 
  as possible substitutes of \bhs, {including gravitational lensing and shadows, possible orbiting 
  star trajectories, accretion disk spectra, and gravitational waves emitted at merges of compact
  objects, of which one or both are \whs.}
  
  Some of the papers evidently go beyond the traditional subjects of \wh\ physics, concerning classical
  traversable \whs. Thus, Sergey Bondarenko \cite{8} considers their quantum counterparts, called 
  quantum \whs, and, in particular, their possible role in the origin of a small cosmological constant
  {favored by observations}.  Elias Zafiris and Albrecht von M\"uller \cite{12} discuss the possible 
  role of Planck-scale \whs\ in the so-called ``ER=EPR'' conjecture in quantum entanglement. Alexander 
  Kirillov and Elena Savelova \cite{13} discuss the conditions under which Planck-scale virtual \whs\ 
  may be converted into macroscopic and observable ones. {The present authors together with
  Pavel Kashargin \cite{10} discuss \whs\ that can emerge in the framework of \GR\ without any exotic 
  matter in nonstatic space-times.}
  
  {The outstanding progress of observational astronomy and cosmology, above all, the discovery of 
  gravitational waves and the pictures obtained by the Event Horizon Telescope, are apparently 
  reflected in the wealth of studies devoted to possible observable effects due to \whs. In the
  present issue, we see the discussions of high-energy particle collisions in \wh\ space-times
  \cite{3}, gravitational lensing by \whs\ with unusual topologies \cite{5}, peculiar features of 
  accretion flows \cite{7} and nearby stellar orbits \cite{9, 14} and possible manifestations of a 
  fractal distribution of primordial \whs\ \cite{8}.}      
  
  It should be noted that, with the whole diversity of \wh\ studies presented in this Special issue, 
  some areas turned out to be almost unmentioned. These are axially and cylindrically symmetric 
  \whs\ with or without rotation (though, some effects of rotation are discussed in \cite{3}) --- 
  so we would here refer to the reviews \cite{cy, ax} and references therein. These are also stability 
  studies concerning all kinds of perturbations of static or stationary \whs\ {(note that Ref.\,\cite{1} 
  only discusses the stability of thin-shell \whs\ with respect to shell motion), so please see, e.g., 
  \cite{stab1, stab2} and references therein for more general studies.}
  
  Concluding, we would like to say that the study of wormholes is far from being complete, and one 
  can expect many new interesting physical and mathematical results in this relevant and
  promising area of physics.

%%%%%%%%%%%%%%%%%%%%%%%%%%%%%%%%%%%%%%%%%%
%\authorcontributions{
%	Individual contributions of authors are the following: The authors have equally contributed to
%	all aspects of the present paper. The authors have read and agreed to the published version of 
%	the manuscript.
%}

%%%%%%%%%%%%%%%%%%%%%%%%%%%%%%%%%%%%%%%%%%
\Funding{S.V.S. is supported by the RSF grant No. 21-12-00130. Partially, this work was done in the framework of the Russian Government Program of Competitive Growth of the Kazan Federal University.
K.B. acknowledges partial support from the Ministry of Science and Higher Education of the Russian Federation, Project ``Fundamental properties of elementary particles and cosmology'' 
No. 0723-2020-0041, and from Project No. FSSF-2023-0003.}

%%%%%%%%%%%%%%%%%%%%%%%%%%%%%%%%%%%%%%%%%%
\ConflictThey

\label{lastpage}

\end{document}